# Third order dielectric time dependent susceptibilities and the "Box Model"

## R. M. Pick[1]

*IMPMC, Université P. et M. Curie et CNRS-UMR 7590, Paris (France)*

We present here the calculation of the "time dependent" moduli of the third order, dielectric, susceptibilities measured at the frequency ω of the applied electric field and at the frequency 3 ω. The calculation is performed within the frame work of the "box model", both in the ideal case of a pure third order polarization and in the practical case studied by Samanta and Richert [J. Chem. Phys. **140**, 247101 (2014)] where appears also a first order polarization contribution. We show that, in the two cases, those two modules have a largely different dynamics and that the results can be easily compared with experiments thanks to the data gathered by Samanta and Richert. This should provide a new test on the validity of the "box model".

The relaxation dynamics, at a given temperature, of glass forming liquids has been the subject of a very large number of theoretical and experimental studies. Already one century ago, it has been proposed [1] that relaxation is an heterogeneous process in the sense that different regions of the liquid relax with a different, Debye, dynamics, each of them being characterized by some relaxation time, $\tau$, the life time of the different regions being independent of $\tau$. The Time Temperature Superposition result [2], generally obtained for this relaxation dynamics, indicate that the probability distribution of those relaxation times is temperature independent when written under the form $P\left(\tau/\tau_\alpha(T)\right)$, $\tau_\alpha(T)$ being a strongly temperature dependent mean relaxation time.

Different techniques have been used [3] to ascertain this heterogeneous picture. In particular, in the specific case of the dielectric properties of these liquids, the Dielectric Hole Burning experiments [4] have provided a positive answer to the heterogeneous hypothesis. In its most recent version, a strong, sinusoidal, electric field, with frequency $\omega/2\pi = 1/T_{mod}$ and amplitude $E_0$, is applied to a capacitor whose cavity is filled with the glass forming liquid. Two routes have been used to exploit this technique.

In one of them [5], one looks at the steady state reached after the application of this strong field, recording over a large frequency range, in the vicinity of the liquid glass transition temperature $T_g$, the part of the polarization which is proportional to $E_0^3$ and which varies either at the frequency $\omega$, or at the frequency $3\omega$. This defines third order susceptibilities, $E_0^3 \chi_3^1(\omega)$ and $E_0^3 \chi_3^3(\omega)$, through their real and imaginary parts and, consequently, through their moduli, $\bar{\chi}_3^1(\omega)$ and $\bar{\chi}_3^3(\omega)$, respectively. The results are usually expressed as a function of $\omega/\omega_{max}$, where $\omega_{max}$ is the frequency at which the imaginary part of the linear dielectric susceptibility has its maximum. One finds that, at both frequencies $\omega$ and $3\omega$, this third order susceptibility has a maximum whose frequency varies with temperature at the same rate as $\omega_{max}$ and whose amplitude increases with decreasing temperature, in a large temperature range above $T_g$ ($\cong 2*10^{-2} < \Delta T/T_g \lesssim 1.5*10^{-1}$, with $T = T_g + \Delta T$, in [6], for instance). This amplitude variation has given an experimental support to the theoretical idea [7] that those third order susceptibilities capture the temperature dependence of the dynamical amorphous order, the latter growing as T decreases. Such behaviour is somewhat reminiscent of that of ordinary spin glasses [8]. Also, a "toy model", [9], based on the idea of asymmetric double wells, already proposed 15 years ago by Wagner and Kleim [10], has been recently proposed to explain the frequency and temperature behavior of $\chi_3^1(\omega)$ and $\chi_3^3(\omega)$. This new model reformulates the former one, including the new concept that, in a molecular glass forming liquid, a dynamical heterogeneity consists of N correlated molecules but has a dipole moment only proportional to $\sqrt{N}$. This "toy model" does predict a temperature variation of the maxima detected in [6] similar to the experimental one, but its very crude character prevents it to be used for the description of the time dependent experiments that represent the second route.

The latter consists in recording the time variation of the third order polarization (or of its time derivative, the current flowing through the capacitor) before reaching the steady state discussed above. The experimental constraints are such that they require that $\omega_{max}/2\pi$ is of the order of 20 Hz and that $\omega >> \omega_{max}$, so that only one temperature and a few discrete frequencies, e. g. $\omega = 25\,n\,\omega_{max}$ where $n$ is an integer with $1 < n < 20$, have been usually studied [11]. Those experiments have been interpreted with the help of the so-called "box-model" [12], the basis of which can be summarized as follows. Each individual dynamical impurity, characterized by its Debye relaxation time $\tau$,

-independently absorbs energy, in the presence of $E_0$, through the Imaginary Part of the liquid linear susceptibility, $\varepsilon''(\omega)$, so that its thermodynamic temperature is raised by $\Delta T_\tau(t)$;

- this increase of $\Delta T_\tau(t)$ is counterbalanced by an energy flow (with the same time rate) from the dynamical impurity to the liquid, considered as a heat bath with an infinite heat capacity; consequently, $\Delta T_\tau(t)$ eventually reaches an equilibrium value;

- the increase in temperature decreases the value of the Debye relaxation time by an amount proportional to $\Delta T_\tau(t)$ without modifying, in first order, the corresponding distribution function;

- this process, when summed over all the relaxation times, results in a time variation of the total polarization of the liquid which is proportional to $E_0^3$ [13].

This polarization change, $\Delta P(t)$, is monitored, in the experiments reported in, e. g., [11a, 11b ], through the intensity, $I(t)$, passing through a resistor in series with the capacitor ($I(t) \propto \Delta \dot{P}(t)$) and the result may be expressed as $\Delta \log(\tan(\delta(t)))$ where $\Delta(\delta(t))$ is the change, between t'=0 and t'=t, of the phase, $\delta$, between $\Delta P(t)$ and the applied electric field $E(t)$, a weak sinusoidal electric field with the same frequency having been applied to the capacitor for $t < 0$.

Richert and his collaborators studied this quantity by recording, with a mesh at least equal to $0.005\,T_{mod}$, both the intensity and the applied electric field [14]. Each signal was then analyzed by computing

$$\chi_s(\omega,t) \equiv \chi'_s(\omega,t) - i\,\chi''_s(\omega,t) = \frac{2i}{E_0^3\,T_{mod}} \int_t^{t+T_{mod}} \dot{S}(t')\exp(-i\,\omega'\,t')\,dt' \quad \text{with } \omega' = \omega \qquad (1)$$

where $S(t)$ is one of the two quantities quoted above, while $t$ was chosen to be of the form $t = t_0 + n\,T_{mod}$, and

$$\delta_S(t) = \text{Arctg}\left(\chi''_s(\omega,t) \Big/ \chi'_s(\omega,t)\right). \qquad (2)$$

In particular, in the cases of Propylene Carbonate (PC) at 166 K and 2-Metyl-TetraHydro-Furan (MTHF) at 96.1 K [11a], Huang and Richert showed that $\varDelta(\delta(t))$, with $\delta(t) = \delta_I(t) - \delta_E(t)$, exhibited the properties predicted by the "box model". In particular, it was found that the dynamics of $\varDelta \log(\tan(\delta(t)))$ was quite slow, displaying a continuous increase towards saturation with a relaxation time of the order of 5 to 10 $T_{mod}$ in the frequency range probed by these authors.

More recently [15], it was pointed out that the variable t appearing in Eq. (1) need not be taken as a discrete quantity with a $T_{mod}$ periodicity. When taken as a continuous variable, $\varDelta \log(\tan(\delta(t)))$ turns out, within the frame work of the "box model", to exhibit, extra oscillations with a periodicity $T_{mod}\Big/2$ (while a naïve interpretation of Eq. (1) would have also suggested the possible detection of additional oscillations with periods $T_{mod}$ and $T_{mod}\Big/3$). These $T_{mod}\Big/2$ oscillations have been indeed found in a recent paper by Samanta and Richert [16], in PC at 166 K, using the same technique of analysis as in [11a] but with a continuous variation of $t$. Those authors simultaneously showed that, with the electric field value, $E_0 = 282\,\text{kV/cm}$, and the 500 Hz frequency they used in their experiment, a small part of the total polarization at frequency $\omega$ was due to a first order contribution. Analyzing their experimental results, we find a relative contribution of the first order polarization with respect to the third order one of the order $1/4$, [17].

The technique used in Eq. (1) needs not be restricted to the pulsation $\omega' = \omega$ of the applied electric field. In the permanent regime, the third order polarization has also a $3\,\omega$ component, and its time dependence before reaching equilibrium can be studied by imposing, in the r. h. s. of Eq. (1), $\omega' = 3\,\omega$, $S(t)$ still standing for the intensity across the sample. One then obtains what can be referred to as the Intensity susceptibility, $\chi_3^3(\omega,t)$, while the corresponding result for $\omega$ can be labeled as $\chi_3^1(\omega,t)$, the index I (for Intensity) being omitted throughout the present paper. Those quantities may be, in particular, studied through their moduli, $\overline{\chi}_3^3(\omega,t)$ and $\overline{\chi}_3^1(\omega,t)$.

The purpose of the present paper is to compare the time dependence of those two moduli for the PC case studied in [16] in the framework of the "box model", the corresponding distribution function for the relaxation times being taken from [18]. I(t) is deduced from the polarization P(t) by performing, first, the analytical derivation of the $\tau$ contribution to this polarization (Eqs. (A-8) and (A3) of [15]), then numerically summing this contribution over $\tau$ and performing the Fourier Transforms indicated in Eq. (1).

Let us start by ignoring the first order polarization contribution detected in [16]. Fig. 1 represents those two susceptibilities ($T_{mod}^{-1} = 500\,Hz$), where $\overline{\chi}_3^3(\omega,t)$ has been multiplied with respect to $\overline{\chi}_3^1(\omega,t)$ by a scaling factor $38.4$ in order that the two susceptibilities have the same t→∞ asymptotic value. This figure shows that:

a) $\overline{\chi}_3^1(\omega,t)$ is a monotonously increasing function of time (except for the weak oscillations with a $T_{mod}/2$ period, similar to the one already detected in [16] for $\Delta\log(\tan(\delta(t)))$, see also b). The relaxation time relative to this increase is $\approx 5\,T_{mod}$ for $0 < t < 10\,T_{mod}$, then increases to $\approx 8\,T_{mod}$ in the vicinity of $20\,T_{mod}$. Conversely, $\overline{\chi}_3^3(\omega,t)$ is essentially time independent, except for oscillations with origins similar to the preceding ones; however, for $\overline{\chi}_3^3(\omega,t)$, there exist two periods, $T_{mod}/2$, as for $\overline{\chi}_3^1(\omega,t)$, but also weaker oscillations with a $T_{mod}/4$ period. Also, their decrease, both for the maxima and the minima of $\overline{\chi}_3^3(\omega,t)$, are similar and much faster than for $\overline{\chi}_3^1(\omega,t)$: this decrease starts with a relaxation time $\approx T_{mod}$ for $0 < t < 2\,T_{mod}$, then rapidly increases to $5\,T_{mod}$ when $t > 3\,T_{mod}$.

b) The two moduli $\overline{\chi}_3^1(\omega,t)$ and $\overline{\chi}_3^3(\omega,t)$ have, in reality, oscillations of similar amplitudes, when measured, for instance at $t \approx T_{mod}$, a fact masked, on Fig. 1, by the large scaling factor ($38.4$) used. Because both susceptibilities have similar time decreases above $t \approx 3\,T_{mod}$, the ratio of the amplitude of their oscillations remains approximately constant in the whole time domain represented on Fig.1

All these results are in line with those computed for MTHF at 96.1 K for a frequency equal to 1000 Hz in [15], though the later correspond, respectively, to the quantities $\chi''^1_3(\omega, t)$ and $\overline{\chi}^3_3(\omega, t)$, for the polarization, and not to $\overline{\chi}^1_3(\omega, t)$ and $\overline{\chi}^3_3(\omega, t)$ for the intensity. Nevertheless, none of them correspond to the situation met in [16] for PC, because of the absence of a first order contribution to the previous calculation. We have thus repeated the whole procedure, starting from a polarization expressed as $0.25\,P_1(t) + P_3(t)$. The results are reported on Fig 2 where, in order for the two susceptibilities to have the same asymptotic values, the scaling factor is now equal to $68.7$. All the findings concerning the time decrease of the susceptibilities, the relative amplitudes of the oscillations or the period for $\overline{\chi}^1_3(\omega, t)$, remain unchanged. The only difference is in the periods appearing in $\overline{\chi}^3_3(\omega, t)$ which are here $T_{mod}$ and $T_{mod}/3$ instead of $T_{mod}/2$ and $T_{mod}/4$ as in Fig. 1. The origin of this new, and quite visible, $T_{mod}/3$ periodicity originates from the $P_1(t)$ contribution. Indeed, this polarization contains a fast decreasing term with no oscillation (see Eq. (A-3) of [15]) for each heterogeneity. The same is true for the corresponding intensity which once multiplied by $\exp(-3i\omega t)$, see Eq. (1), yields this $T_{mod}/3$ periodicity in the real and imaginary parts of this new susceptibility. Nevertheless, the maximum and the two additional minima, all visible on Fig. 2, decrease in the same manner as the maxima and minima relative to $\overline{\chi}^3_3(\omega, t)$ in Fig. 1.

The present results turn out to disagree with a recent experimental finding of R. Richert [18], who has written, p. 5, end of the second paragraph :"Reanalyzing the data (*of [11a] and [11b]*) has revealed that the moduli and the phases of both $\chi^1_3(\omega, t)$ and $\chi^3_3(\omega, t)$ are subject to similarly slow evolutions in time". The raw data on $I(t)$ obtained in [16] can be analyzed using Eq. (1) with $\omega' = \omega$ and with $\omega' = 3\,\omega$. This would provide the time evolution of those two moduli while the results shown in Fig. 2 indicate that the corresponding oscillations with frequencies $T_{mod}$ and $T_{mod}/3$ should be as visible as the $T_{mod}/2$ oscillations detected in [16]. A similar time evolution of the two susceptibilities, as proposed in [18], would make fortuitous the general and impressive agreement found in [11b, 14, 16] between the "box model" and the experiments on $\chi_3^1(\omega, t)$, calling for a novel model to explain them. The calculations presented here make it clear that such an analysis is possible and thus worthwhile.

# Figures

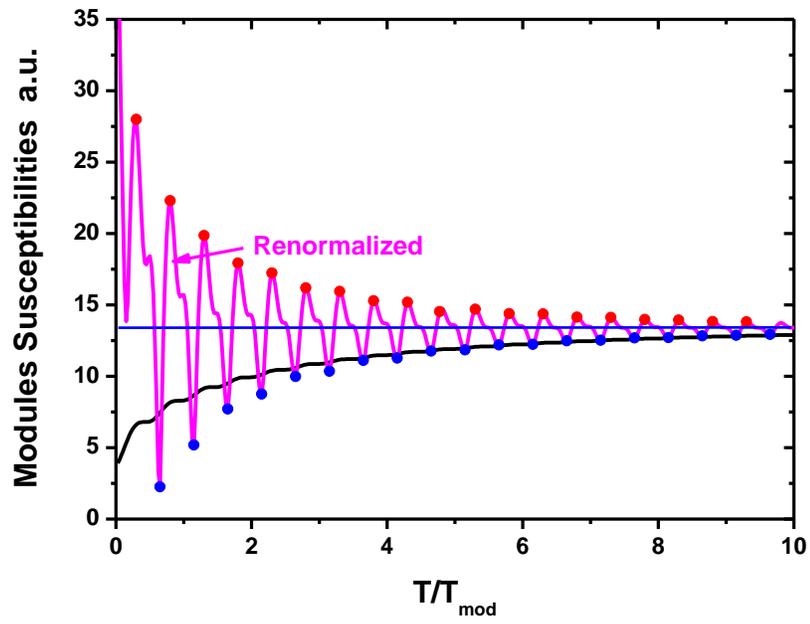

Fig. 1

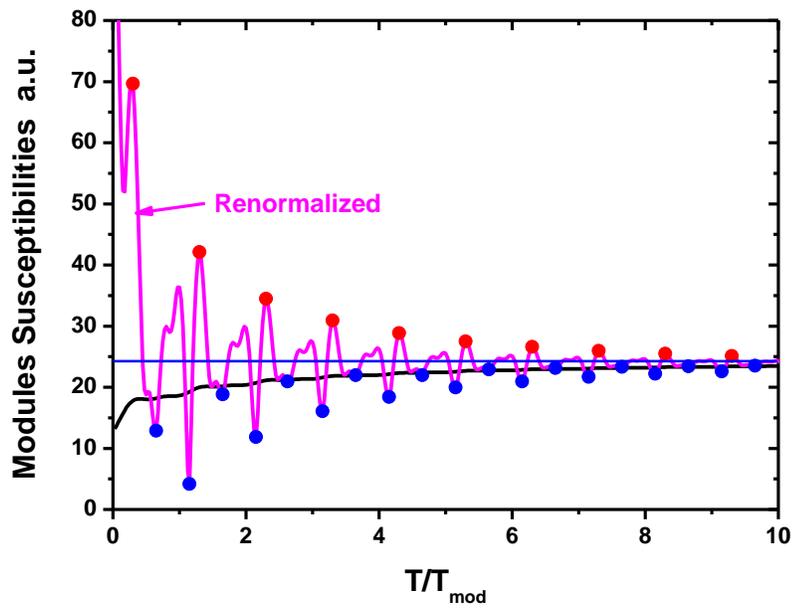

Fig. 2

**Figures Captions**

Fig. 1: $\bar{\chi}_3^1(\omega, t)$, black, and $\bar{\chi}_3^3(\omega, t)$ (multiplied by a factor $38.4$), magenta, for PC at 166 K and an electric field frequency equal to 500 Hz. The maxima and minima of the oscillations of $\bar{\chi}_3^3(\omega, t)$ are represented respectively by red and blue dots.

Fig. 2: Same as on Fig. 1 but for the intensity resulting from a total polarization equal to $0.25\, P_1(t) + P_3(t)$, in accordance with the experimental results of [16]. Note the disappearance of the $T_{mod}/2$ and $T_{mod}/4$ periodicities, replaced by the two $T_{mod}$ and $T_{mod}/3$ ones in $\bar{\chi}_3^3(\omega, t)$ .